\documentclass[11pt]{article}
\usepackage{amssymb}
\usepackage{epsfig}
\begin{document}
\begin{titlepage}

\title{Particle Physics in Intense Electromagnetic Fields
{\footnote{Published in: IL Nuovo Cimento A, 1999, vol.112A,
p.977-1000.}}}

\author{Alexander V. Kurilin{\footnote{E-mail address:
alexkurilin@mtu-net.ru}}\\ Department of Physics\\ Moscow State
Open Pedagogical University\\ Verkhnyaya Radishchevskaya
16-18,\\ Moscow 109004, Russia\\}

\date{Received 3 March 1999}
\maketitle

\abstract
{\normalsize The quantum field theory in the presence of
classical background electromagnetic fields is reviewed.
We give a pedagogical introduction to the Feynman-Furry
method of describing non-perturbative interactions with very
strong electromagnetic fields. A particular emphasis is given
to the case of the plane-wave electromagnetic field for which the
charged particles' wave functions and propagators are presented.
Some general features of quantum processes proceeding in the
intense electromagnetic background are argued. We also discuss
the possibilities of searching new physics through the
investigations of quantum phenomena induced by the strong
electromagnetic environment.

PACS 13.40 - Electromagnetic processes and properties.

PACS 13.40.Hq - Electromagnetic decays.

PACS 13.10 - Weak and electromagnetic interactions of leptons.

PACS 11.15.Tk - Other nonperturbative techniques.}
\end{titlepage}

\section{Introduction}
Quantum field theory in the presence of an intense electromagnetic
background has a long fruitful history. The interest promoting
investigations in this area is based on several reasons. The first
one being purely theoretical is associated with the self-consistency problem of the relativistic quantum theory at very
high energies. It is well known that in quantum electrodynamics
(QED) perturbative vacuum probably is not a true ground state of
the theory because of the fictitious pole in photon propagator in
the ultraviolet region : $\vert k^2\vert \sim
m_e^2\exp(3\pi/\alpha)$ \cite{Moscow_0}. This challenge has
inspired many studies dealing with QED of intense fields in which
the convergence of perturbation series in fine-structure constant
$\alpha=e^2/4\pi$ has been analyzed (see e.g. \cite{FIAN-111},
\cite{FIAN-168}, \cite{FIAN-192} and references therein). The
matter is that ultraviolet behaviour of the theory and the
structure of divergencies determine in a uniform way the
properties of Green functions in very intense electromagnetic
fields and at large momentum transfer. In other words the physics
of extremely strong background fields is intimately connected with
the one at small distances \cite{Migdal-72},\cite{Ritus-77}. This
can be illustrated by the effect of vacuum polarization in QED
which modifies the Lagrange density of Maxwell electrodynamics.
For weak fields one loop corrections have been calculated long ago
by Heisenberg and Euler \cite{Heisenberg-Euler}.
\begin{equation}
{\cal L}=-{\cal F}+\frac{2\alpha^2}{45 m_e^4}
\bigl( 4{\cal F}^2+7{\cal G}^2 \bigr),
\end{equation}
where
\begin{equation}
\label{invariants}
{\cal F}={1\over 4}F_{\mu\nu} F^{\mu\nu}={1\over 2}
\bigl({\bf B}^2-{\bf E}^2\bigr),
\qquad
{\cal G}=-{1\over 4}F_{\mu\nu}\tilde F^{\mu\nu}={\bf E B}.
\end{equation}
However the above expression is valid only if the field strength
is relatively small (${\cal F, G}\ll m_e^4/e^2$). The domain of
extremely strong fields (${\cal F}\gg m_e^4/e^2; {\cal G}\ll
{\cal F}$) is described by another formula
\cite{Schwinger-51} which in logarithmic approximation can be
written as \cite{L-eff}
\begin{equation}
\label{L-eff}
{\cal L}=-{\cal F} + \frac{\alpha}{6\pi}{\cal F}
\ln\Bigl( \frac{2 e^2{\cal F}}{m_e^4}\Bigr)
\end{equation}
We see that asymptotically the effective Lagrangian
(\ref{L-eff}) vanishes as far as the field strength
$B=\sqrt{2{\cal F}}$  tends to the critical value $\sim m_e^2/e
\exp(3\pi/\alpha)$. So, the Moscow zero-charge problem
\cite{Moscow_0} reveals itself as a nought of effective
Lagrangian if we impose the following interrelation:
\begin{equation}
\label{k-B}
\vert k^2\vert\sim eB=e\sqrt{2{\cal F}}
\end{equation}
Two loop calculations \cite{Ritus-75} have confirmed that the
analogy (\ref{k-B}) is retained in all orders of perturbation.
From this viewpoint the physics of intense electromagnetic fields
is a powerful theoretical tool for searching the stable
non-perturbative vacuum state in which the artifacts mentioned
above can be removed.

Let us also note, that the problem of vacuum instability and phase
transitions in a strong background field has received additional
interest in the context of non-Abelian gauge theories
\cite{vacuum-B}. These investigations rely essentially on the
philosophy and technique elaborated in the framework of QED with
an intense background.

The second motivation for studying particles interactions with
very strong electromagnetic fields originates from the quantum
theory of synchrotron radiation \cite{synchrotron}. The process of
bremsstrahlung by electrons in a homogeneous magnetic field is has
long been known. Classical electrodynamics provides a satisfactory
description of the radiation power and spectrum being observed in
synchrotrons (for a recent review see \cite{Ternov-95}). At the
same time particular phenomena require the apparatus of QED with
an intense magnetic field to be applied for an adequate
theoretical explanation. Systematic investigations of quantum
processes in the presence of strong magnetic fields have been
initiated in works of Sokolov and Ternov \cite{Sokolov-Ternov} and
were continued by their collaborators \cite{MSU}. These
investigations resulted in discovering many interesting
theoretical predictions some of which have been confirmed
experimentally. Here we can highlight the famous effect of
self-polarization of electrons in storage rings due to the
synchrotron radiation.

The third reason stimulating interest in physics with intense
electromagnetic background is caused by recent CERN experiments
involving SPS accelerator in which the particle interactions with
single crystals were explored \cite{CERN-exp1}, \cite{CERN-exp2}.
Notice that, because of regular and systematic structure,
electromagnetic interactions in single crystals are substantially
enhanced in comparison with the amorphous medium. This gives a
unique possibility of creating a strong electromagnetic background
for relativistic particles penetrating single crystal near axial
or planar directions. Thus we have an opportunity for searching
new physics in such unusual conditions because quantum phenomena
in the presence of the intense electromagnetic field differ
considerably from those occurring in a vacuum. Due to non-linear
and non-perturbative influence of the external field one can
observe absolutely new processes which as a rule are forbidden
under normal conditions. The electromagnetic field open new
channels of reactions taking away the bans for transitions
between definite quantum states. The most wonderful feature is the
possibility for very light particles to decay into heavy species
capturing a lacking amount of energy from the intense
electromagnetic environment. It would be pertinent to mention here
the process of electron-positron pairs production by solitary
photons incident along crystal axes. The data of experiments CERN
WA-81, CERN NA-046 \cite{CERN-exp2} concerning with this reaction
$\gamma\rightarrow e^+e^-$ are in a satisfactory agreement with
predictions of QED with the strong electromagnetic field
\cite{Baier-UFN} (see also \cite{crystals}).

Thus, we see that particle physics in the presence of
non-perturbative background fields is an impetuously developing
branch of physical sciences having many interesting
applications. Unfortunately it is impossible to itemize all
exploitations of this theoretical technique because they are
very numerous. Here I can only mention some other important
employments such as the problem of vacuum in quantum
chromodynamics \cite{QCD}, astrophysics of neutron stars
\cite{astrophysics}, the cosmology of the early Universe
\cite{cosmology} and etc.

Lastly, let us say some words about the method itself. The
philosophy of calculations in strong electromagnetic fields
originates from the Feynman non-perturbative approach in QED
\cite{Feynman} which was elaborated in details by Furry
\cite{Furry}. This technique has become widely known as the
``Furry picture''. Classical examples of QED with external fields
were given in works of Schwinger \cite{Schwinger-51},
\cite{Schwinger}. Further developments of the Feynman-Furry method
were achieved in the research of Fradkin and Gitman with their
collaborators who have generalized and modified the above approach
for a wide class of electromagnetic fields with unstable vacuum
(see \cite{Fradkin-Gitman} and references therein). It is worth
noting that the procedure of second quantization in the intense
background field is a separate, uneasy task having common features
with the quantum field theory in a curved space-time
\cite{gravity}. A particular difficulty is connected with
appropriate treatment of the classical solutions of relativistic
wave equations that do not admit one particle interpretation. For
instance, in a case of a constant uniform electric field exact
solutions of the Dirac equation have a form which cannot be
splitted into positive and negative frequency modes corresponding
to particle and antiparticle wave functions. The later
circumstance is regarded as the instability of the physical vacuum
with electric fields in relation with the processes of pairs
production. However, as was shown by Schwinger \cite{Schwinger-51}
the probability of pairs creation from the vacuum becomes feasible
only in very strong electric fields being comparable with the
critical value $E_{cr}=m_e^2/e= 1.32\cdot 10^{18} V/m$.

Inasmuch as present values of the field strength being available
in experiments are much less than $E_{cr}$ it is advantageous to
employ an effective approximation which was proposed in works of
Ritus and Nikishov \cite{Ritus}. The main idea comes from a
detailed analysis of probabilities of quantum phenomena in an
arbitrary constant homogeneous electromagnetic field. Suppose
that the initial quantum state is characterized by the only one
particle with momentum $p^\mu$. Then, in general the probability
of quantum transition to some final state $P_{fi}$ depends on
the background field strength $F_{\mu\nu}$ through the following
Lorentz-invariant dimensionless parameters:
\begin{equation}
\label{parameters}
a=\frac{e^2{\cal G}}{m_e^4}=\frac{e^2}{m_e^4}{\bf E B},
\qquad
b=\frac{e^2{\cal F}}{m_e^4}=\frac{e^2}{2 m_e^4}
\bigl({\bf B}^2-{\bf E}^2\bigr).
\end{equation}
\begin{equation}
\label{chi}
\chi=\frac{e}{m_e^3}\sqrt{-(F_{\mu\nu} p^\nu)^2}=
\frac{e}{m_e^3}\bigl[ (p_0 {\bf E} + {\bf p\times B})^2
-({\bf p E})^2 \bigr]^{1/2}
\end{equation}
The domain of electromagnetic fields that currently can be
handled in a laboratory satisfies the conditions: $a\ll 1,
b\ll 1$ (i.e. $E\ll E_{cr}, B\ll B_{cr}=m_e^2/e=4.41
\cdot 10^9 T$). Now, if the initial particle is relativistic
($p_0\gg m_e$) then $\chi^2\gg a; b$ and one can
expand the probability $P_{fi}$ into a series of these small
quantities:
\begin{equation}
\label{approximation}
P_{fi}(\chi, a, b)=P(\chi, 0, 0) + a \frac{\partial P}{\partial a}
(\chi, 0, 0)+ b \frac{\partial P}{\partial b}(\chi, 0, 0) + \ldots
\end{equation}
The first term in eq.(\ref{approximation}) represents the
probability of quantum transition in a so-called crossed field,
which is a combination of the orthogonal electric and magnetic
fields having the same magnitude (${\cal F}={\cal G}=0$). So, the
formula describing a reaction in the crossed field provides a good
approximation for that in an arbitrary constant electromagnetic
field if one can omit in eq.(\ref{approximation}) the corrections
proportional to small parameters $a, b$. The merits of the crossed
field approximation lie in the most simple form of
charged-particle wave functions and propagators which greatly
facilitate the scheme of calculations. All what said above may be
summarized as the following prescription: if you want to
investigate a process in a constant homogeneous electromagnetic
field, then the most straightforward way is to calculate  the
corresponding probability assuming that ${\cal F}={\cal G}=0$
(\ref{invariants}). The obtained result depending on the sole
background field parameter $\chi$ (\ref{chi}) will simulate the
exact one as the first term in the expansion
(\ref{approximation}). Note, that the crossed field is a
particular case of the plane-wave electromagnetic field fitting
the choice: $A_\mu(\varphi)=a_\mu\varphi;\quad F_{\mu\nu}=k_\mu
a_\nu - k_\nu a_\mu$ (see below).

The purpose of this paper is to give a pedagogical introduction to
the particle physics in the presence of an intense electromagnetic
background. I briefly reproduce the main milestones of the
Feynman-Furry method allowing for non-perturbative interactions
with very strong electromagnetic fields. Taking into account the
spectrum of the known electrically charged particles, I consider
the cases of scalar, spinor and vector quantum fields separately
and obtain various forms for wave functions and propagators that
are suitable for practical calculations (sections 3,4,5). Then I
discuss some general features that are typical for quantum
phenomena in the background electromagnetic field (section 6). And
finally, I analyze the possibilities of searching new physics in
the framework of Minimal Standard Model through the investigation
of quantum effects induced by the strong electromagnetic
background.

\section{Scalar Particles in an Intense Electromagnetic Background}
Let us consider the Lagrangian of scalar particles with charge
$e>0$ interacting with a strong electromagnetic field.
\begin{equation}
\label{Lag}
{\cal L}=(\partial_\mu\phi^+)(\partial^\mu\phi^-)-m^2\phi^+
\phi^- +ieA^\mu(\phi^+\partial_\mu\phi^- -\phi^-
\partial_\mu\phi^+)+e^2\phi^+\phi^-(A^\mu A_\mu)
\end{equation}
The electromagnetic potential $A_\mu(x)$ is regarded here as a
given function of space-time coordinates which corresponds to
the classical field strength tensor $F_{\mu\nu}=\partial_\mu
A_\nu - \partial_\nu A_\mu$. We wish to describe the impact of
the external electromagnetic field in a non-perturbative manner
employing exact solutions of the equation of motion which
follows from the Lagrangian (\ref{Lag}).
\begin{equation}
\label{Klein}
\Bigl[(\partial_\mu\pm ieA_\mu)^2 + m^2 \Bigr]\phi^\pm (x) = 0
\end{equation}
The greatest embarrassment of this approach is that of solving
analytically equation (\ref{Klein}) and to obtain non-perturbative
wave functions for an arbitrary predetermined configuration of the
background field. At present exact solutions are known for the
case of a constant uniform electromagnetic field, for the Coulomb
field potential and for the so-called Redmond's configuration (see
e.g. \cite{Bagrov}). Besides there is an explicit expression for
the scalar particle wave function in the presence of a plane
electromagnetic wave, which is characterized by some unspecified
potential $A_\mu(\varphi)$ depending on space-time coordinates
$x^\mu$ only through the phase $\varphi=k_\mu x^\mu=\omega t-{\bf
k x}$. This is the case we intend to discuss in more details.

The starting point of our analysis is to invoke the WKB method
for the equation (\ref{Klein}) which as it comes out will give
us the exact result in the first approximation. Now, substituting
$\phi^+(x)=e^{iS(x)}$ and imposing the Lorentz gauge for the
vector potential, we get
\begin{equation}
\label{Hamilton}
(\partial_\mu S + eA_\mu)^2-m^2=i\partial_\mu\partial^\mu S.
\end{equation}
In the semiclassical limit being considered the right-hand side of
eq. (\ref{Hamilton}) can be neglected and it gives us the
Hamilton-Yacobi equation for the classical action of a
relativistic particle moving in a plane-wave electromagnetic
field. The classical action as a function of coordinates $x^\mu$
and the vector of initial momentum $p^\mu$ is well known and it
can be written in the following form:
\begin{equation}
\label{action}
S(x,p)=-p_\mu x^\mu - \int\limits_0^{kx}\Biggl(\frac{e p_\mu
A^\mu(\varphi)}{p_\nu k^\nu}-\frac{e^2 A_\mu(\varphi)
A^\mu(\varphi)}{2 p_\nu k^\nu}\Biggr)d\varphi + S_0.
\end{equation}
Here it is implied that the wave vector $k^\mu$ obeys the
isotropic and transversal conditions
\begin{equation}
\label{k-wave}
k^2=0, \qquad\qquad k^\mu A_\mu=0.
\end{equation}
In order to obtain quantum corrections one must insert formula
(\ref{action}) into eq.(\ref{Hamilton}) and take into account
the right-hand side of this expression. However it is easy to
see that classical action (\ref{action}) makes both sides of eq.
(\ref{Hamilton}) vanish if the momentum vector $p^\mu$ is taken
on the mass shell: $p_0=\sqrt{{\bf p}^2+m^2}$. So, we have found
an exact classical solution of eq.(\ref{Klein}) which should be
treated as the scalar particle wave function in a background
electromagnetic wave field. For practical use the wave function
must be normalized in a proper way. Assuming that the
configuration volume space is equal to $V$, we finally get
\begin{equation}
\label{scalar-wf}
\phi^\pm(x,{\bf p})=\frac{1}{\sqrt{2p_0 V}}\exp\biggl[-ipx-i\int
\limits_0^{kx}\biggl(\pm\frac{epA}{pk}-\frac{e^2 A^2}{2pk}
\biggr)d\varphi\biggr].
\end{equation}

The next step is to realize the second quantization procedure. As
usually, we decompose the scalar field into a series of the
obtained-above solutions involving the Fock creation and
annihilation operators.
\begin{equation}
\label{quant-scalar+}
\phi^+(x)=\sum_{\bf p}\Bigl[a_{\bf p}\phi^+(x,{\bf p})+
b_{\bf p}^{\dagger}\phi^{-*}(x,{\bf p})\Bigr]
\end{equation}
\begin{equation}
\label{quant-scalar-}
\phi^-(x)=\sum_{\bf p}\Bigl[a_{\bf p}^{\dagger}\phi^{+*}(x,{\bf p})+
b_{\bf p}\phi^-(x,{\bf p})\Bigr]
\end{equation}
These operators must satisfy a standard set of commutation
relations
\begin{eqnarray}
\label{commutators-1}
\bigl[a_{\bf p};a_{\bf p'}^{\dagger}\bigr] = \delta_{\bf p p'},\qquad
\bigl[b_{\bf p};b_{\bf p'}^{\dagger}\bigr] = \delta_{\bf p p'},\nonumber\\
\bigl[a_{\bf p};a_{\bf p'}\bigr] = \bigl[b_{\bf p};b_{\bf p'}\bigr]=
\bigl[a_{\bf p};b_{\bf p'}\bigr] = \bigl[a_{\bf p};b_{\bf p'}^{\dagger}
\bigr]=0,
\end{eqnarray}
which enable us to reproduce the ordinary scheme of quantum field
theory. All further reasoning almost exactly repeats the ones of
the conventional approach. The negative frequency modes of
eq.(\ref{Klein}) are regarded as the scalar antiparticles with
wave functions $\phi^-(x,{\bf p})$ and the opposite charge $-e<0$.
The physical vacuum is defined as a ground quantum state with no
particles present: $a_{\bf p}\vert 0> =b_{\bf p}\vert 0>=0$, and
the Feynman propagator is determined by the formula
\begin{equation}
\label{propagator-1s}
D(x,x')=i<0\vert T\phi^+(x)\phi^-(x')\vert 0>
\end{equation}
So, the essence of the Furry picture of interaction with a
background electromagnetic field is to replace the ordinary de
Broglie's wave functions by some exact solutions which incorporate
adequately all non-perturbative effects. As a calculation
prescription, we present explicitly the modification of the scalar
particle propagator (\ref{propagator-1s}) in a plane-wave field:
\begin{equation}
\label{propagator-2s}
D(x,x')=\int\frac{d^4p}{(2\pi)^4}\frac{1}{m^2-p^2-i0}
\exp\biggl[-ip(x-x')-i\int\limits_{kx'}^{kx}
\biggl(\frac{epA}{pk}-\frac{e^2 A^2}{2pk}\biggr)d\varphi\biggr].
\end{equation}
Notice that the propagator (\ref{propagator-2s}) is a Green
function of the Klein-Gordon equation (\ref{Klein}) id est it
obeys inhomogeneous equation with the Dirac $\delta$-function in
the right-hand side:
\begin{equation}
\label{Green-1}
\Bigl[(\partial_\mu+ieA_\mu)^2+m^2\Bigr]D(x,x')=\delta^4(x-x')
\end{equation}
Another useful representation for the scalar propagator can be
obtained from eq.(\ref{Green-1}) through the Fock-Schwinger proper
time formalism \cite{Schwinger-51}, \cite{Fock}. A comprehensive
expounding of this method can be found in many text books
\cite{Itzykson-Zuber}. For the sake of completeness let us
introduce here the final expression for the propagator
(\ref{propagator-1s}) in such an approach which is equivalent to
eq.(\ref{propagator-2s}):
\begin{eqnarray}
\label{propagator-3s}
D(x,x')=\frac{\Phi(x,x')}{(4\pi)^2}\int\limits_0^\infty
\frac{ds}{s^2}\exp\biggl[-ism^2-i\frac{(x-x')^2}{4s}
+\frac{ie^2s}{(kx-kx')}\int\limits_{kx'}^{kx}A_\mu(\varphi)
A^\mu(\varphi)d\varphi-\nonumber\\
-\frac{ie^2s}{(kx-kx')^2}\int\limits_{kx'}^{kx}
d\varphi_1\int\limits_{kx'}
^{kx}d\varphi_2A_\mu(\varphi_1)A^\mu(\varphi_2)\biggr],
\end{eqnarray}
where the gauge factor $\Phi(x,x')$ is determined by the
integral along the straight line from the point $x_\mu$ to the
point $x'_\mu$:
\begin{equation}
\label{phase}
\Phi(x,x')=\exp\biggl[-ie\int\limits_{x'}^x A_\mu(z)dz^\mu\biggr]=
\exp\biggl[-ie\frac{(x-x')^\mu}{(kx-kx')}\int\limits_{kx'}^{kx}
A_\mu(\varphi)d\varphi\biggr].
\end{equation}

\section{Charged Fermions in an Intense Electromagnetic Field}
The results of previous section can be easily generalized for
the spin $1/2$ particles being governed by QED-like Lagrangian
\begin{equation}
\label{QED}
{\cal L}={i\over 2}\overline{\psi}\gamma^\mu(\partial_\mu\psi)-
{i\over 2}(\partial_\mu\overline{\psi})\gamma^\mu\psi-m
(\overline{\psi}\psi)
-eA_\mu (\overline{\psi}\gamma^\mu\psi),
\end{equation}
which gives the Dirac equation for the four component bispinor $\psi$:
\begin{equation}
\label{Dirac}
\Bigl(i\gamma^\mu\partial_\mu-e\gamma^\mu A_\mu-m\Bigr)\psi(x)=0
\end{equation}
Exact solutions in a plane electromagnetic wave have been
obtained by Volkov \cite{Volkov} and they are deduced from the
Klein-Gordon equation (\ref{Klein}) by a standard substitution
(see also \cite{Berestetskii}):
\begin{equation}
\label{subst}
\psi(x)=\Bigl(i\gamma^\mu\partial_\mu-e\gamma^\mu A_\mu+m\Bigr)
\varphi(x).
\end{equation}
The squared Dirac equation
\begin{equation}
\label{2-Dirac}
\Bigl[(\partial_\mu + ieA_\mu)^2+m^2+{e\over 2}\sigma^{\mu\nu}
F_{\mu\nu}\Bigr]\varphi(x) = 0
\end{equation}
differs from that for scalar particles only in the last term
arising because of $\gamma$-matrices commutator
\begin{equation}
\sigma^{\mu\nu}={i\over 2}(\gamma^\mu\gamma^\nu-
\gamma^\nu\gamma^\mu).
\end{equation}
Note that in a plane electromagnetic wave all powers of this
term exceeding the first are equal  to zero due to
eq.(\ref{k-wave}). So the desired solution of equation
(\ref{2-Dirac}) is a slightly modified form of expression
(\ref{scalar-wf}).
\begin{equation}
\label{solution}
\varphi(x)=\exp\biggl[-ipx-i\int\limits_0^{kx}\biggl(\frac{epA}{pk}
-\frac{e^2
A^2}{2pk}\biggr)d\varphi\biggr]\biggl[1+\frac{e}{2pk}\hat k\hat A(kx)
\biggr]v,
\end{equation}
where$$ \hat k=k_\mu\gamma^\mu, \qquad\qquad \hat A=\gamma^\mu
A_\mu, $$ and $v$ is an arbitrary constant bispinor. Now,
returning to eq.(\ref{subst}), we derive the sought for
non-perturbative wave function of a charged fermion in an intense
electromagnetic background:
\begin{equation}
\label{spinor-wf}
\Psi^\pm(x,{\bf p})=\exp\biggl[-ipx-i\int\limits_0^{kx}
\biggl(\pm\frac{epA}{pk}-\frac{e^2 A^2}{2pk}\biggr)d\varphi\biggr]
\biggl[1\pm\frac{e}{2pk}\hat k\hat A(kx)\biggr]\frac{u_\sigma (p)}
{\sqrt{2p_0 V}}
\end{equation}
Polarization effects are described here by an ordinary Dirac
bispinor $$u_{\sigma}(p)=\sqrt{2p_0 V}(\gamma^\mu p_\mu+m)v$$ which is
subject to the following restrictions:
\begin{equation}
\label{bispinor}
(\gamma^\mu p_\mu-m)u_{\sigma}(p)=0 \qquad
\bar u_{\sigma}(p)u_{\sigma}(p)=2m \qquad
\bar u_{\sigma}(p)\gamma^\mu u_{\sigma}(p)=2p^\mu
\end{equation}
The second quantization is carried out by analogy with eq.
(\ref{quant-scalar+}),(\ref{quant-scalar-})
\begin{equation}
\label{quant-fermion}
\psi(x)=\sum_{{\bf p}\sigma}\Bigl[a_{{\bf p}\sigma}\Psi^+(x,{\bf p})+
b_{{\bf p}\sigma}^{\dagger}\Psi^{-c}(x,{\bf p})\Bigr]
\end{equation}
\begin{equation}
\label{quant-antifermion}
\overline{\psi}(x)=\sum_{{\bf p}\sigma}\Bigl[a_{{\bf p}\sigma}^{\dagger}
\overline{\Psi^+}(x,{\bf p})+b_{{\bf p}\sigma}
\overline{\Psi^{-c}}(x,{\bf p})\Bigr]
\end{equation}
and involves also the antifermion wave functions being obtained
through the $C$-matrix of charge conjugation:
\begin{eqnarray}
\label{C-function}
\Psi^{-c}(x,{\bf p})=C\overline{\Psi}^{-T}(x,{\bf p})=
-\gamma^0 C\Psi^{-*}(x,{\bf p})\nonumber\\
\overline{\Psi^{-c}}(x,{\bf p})=-\Psi^{-T}(x,{\bf p})C^{\dagger}
\end{eqnarray}
A specific form of this matrix is inessential provided that it
satisfies the following properties:
\begin{equation}
C^{\dagger}=C^{-1} \qquad C^T=-C \qquad
C^{-1}\gamma^\mu C=-\gamma^{\mu T}
\end{equation}
For example we can choose $C=-i\gamma^0 \gamma^2$ in Weyl and
Dirac $\gamma$-matrices representations.

Taking into account the spin-statistics relation, we must impose
the anticommutation rules for the Fock operators
$a_{{\bf p}\sigma}, b_{{\bf p}\sigma}$ which have no difference
with the conventional ones:
\begin{eqnarray}
\label{commutators-2}
\{ a_{{\bf p}\sigma};a_{{\bf p'}\sigma'}^{\dagger} \}=
\delta_{\bf p p'}\delta_{\sigma\sigma'} \qquad
\{ b_{{\bf p}\sigma};b_{{\bf p'}\sigma'}^{\dagger} \}=
\delta_{\bf p p'}\delta_{\sigma\sigma'} \nonumber\\
\{ a_{{\bf p}\sigma};a_{{\bf p'}\sigma'} \}=
\{ b_{{\bf p}\sigma};b_{{\bf p'}\sigma'} \}=
\{ a_{{\bf p}\sigma};b_{{\bf p'}\sigma'} \}=
\{ a_{{\bf p}\sigma};b_{{\bf p'}\sigma'}^{\dagger} \}=0
\end{eqnarray}
The corresponding Feynman propagator is defined as usual
\begin{equation}
\label{propagator-1f}
S(x,x')=-i<0\vert T\psi(x)\overline{\psi}(x')\vert 0>
\end{equation}
and has the most simple form in the basis of exact solutions
(\ref{spinor-wf}):
\begin{eqnarray}
\label{propagator-2f}
S(x,x')=\int\frac{d^4p}{(2\pi)^4}
\biggl[1+\frac{e\hat k}{2pk}\hat A(kx)\biggr]
\frac{\gamma^\mu p_\mu + m}{p^2-m^2+i0}
\biggl[1-\frac{e\hat k}{2pk}\hat A(kx')\biggr]\cdot\nonumber\\
\cdot\exp\biggl[-ip(x-x')-i\int\limits_{kx'}^{kx}
\biggl(\frac{epA}{pk}-\frac{e^2 A^2}{2pk}\biggr)d\varphi\biggr].
\end{eqnarray}
It is often more convenient to use the Fock-Schwinger
representation which enables us to rewrite the last formula as
follows.
\begin{eqnarray}
\label{propagator-3f}
S(x,x')=-\frac{\Phi(x,x')}{(4\pi)^2}\int\limits_0^\infty
\frac{ds}{s^2}\exp\biggl[-ism^2-i\frac{(x-x')^2}{4s}
+\frac{ie^2s}{(kx-kx')}\int\limits_{kx'}^{kx}A_\mu(\varphi)
A^\mu(\varphi)d\varphi-
\nonumber\\
-\frac{ie^2s}{(kx-kx')^2}\int\limits_{kx'}^{kx}
d\varphi_1\int\limits_{kx'}^{kx}d\varphi_2 A_\mu(\varphi_1)
A^\mu(\varphi_2)\biggr]
\Biggl\{m+\frac{(\hat x -\hat x')}{2s}+ m e s \hat k
\frac{\hat A (kx)-\hat A (kx')}{(kx-kx')}+
\nonumber\\
+\frac{e}{2(kx-kx')}\Bigl[\hat k\hat A(kx)(\hat x -\hat x')
-(\hat x - \hat x')\hat k\hat A(kx')\Bigr]
+\frac{e}{(kx-kx')}\int\limits_{kx'}^{kx}\hat A(\varphi)d\varphi-
\nonumber\\
-e\hat k\frac{(x-x)^\mu}{(kx-kx')^2}\int\limits_{kx'}^{kx}
A_\mu(\varphi)d\varphi-\frac{e^2 s \hat k}{(kx-kx')}
\hat A(kx) \hat A(kx')
+\frac{e^2 s \hat k}{(kx-kx')^2}\int\limits_{kx'}^{kx}
A_\mu(\varphi) A^\mu(\varphi)d\varphi+
\nonumber\\
+\frac{e^2 s \hat k}{(kx-kx')^2}\int\limits_{kx'}^{kx}
\Bigl[\hat A(kx) \hat A(\varphi)+\hat A(\varphi) \hat A(kx')
\Bigr] d\varphi
-\frac{2 e^2 s \hat k}{(kx-kx')^3}\int\limits_{kx'}^{kx}
d \varphi_1 \int\limits_{kx'}^{kx} d \varphi_2
A_\mu(\varphi_1) A^\mu(\varphi_2) \Biggr\}
\end{eqnarray}
One can easily check that fermion propagator
(\ref{propagator-2f}), (\ref{propagator-3f}) is a Green function
of the Dirac equation (\ref{Dirac}) in the presence of the
intense electromagnetic field.
\begin{equation}
[i\gamma^\mu \partial_\mu - e\gamma^\mu A_\mu - m]
S(x,x')=\delta^4(x-x').
\end{equation}

\section{W-bosons in a Background Electromagnetic Field}
Let us now concentrate on electromagnetic interactions of weak
intermediate vector bosons $W^{\pm}$. The corresponding part of
the Lagrangian of the Glashow-Weinberg-Salam model has the form
(see e.g. \cite{GWS})
\begin{eqnarray}
\label{L-W}
{\cal L}=-{1\over 2}(\partial_\mu W_\nu^+ -\partial_\nu W_\mu^+)
(\partial^\mu W^{-\nu} -\partial^\nu W^{-\mu}) + m_W^2 W^+_\mu
W^{-\mu} -ie F^{\mu\nu}W^+_\mu W^-_\nu - \nonumber\\
-ie(\partial_\mu W_\nu^+ -\partial_\nu W_\mu^+) W^{-\mu} A^\nu
+ie(\partial_\mu W_\nu^- -\partial_\nu W_\mu^-) W^{+\mu} A^\nu -
e^2(A_\mu A^\mu)(W^+_\nu W^{-\nu})+\nonumber\\
+ e^2(A^\mu W^+_\mu)(A^\nu W^-_\nu)
-{1\over \xi}(\partial^\mu W^+_\mu +ieA^\mu W^+_\mu)
(\partial^\nu W^-_\nu -ieA^\nu W^-_\nu)
\end{eqnarray}
The last summand in (\ref{L-W}) arises from a non-linear gauge
fixing term
\begin{eqnarray}
{\cal L}_{gf}=-{1\over 2\xi}
(\partial^\mu A_\mu^1-g A_\mu^2 A^{3\ \mu} +\xi m_W \varphi_1)^2
- {1\over 2\xi}
(\partial^\mu A_\mu^2+g A_\mu^1 A^{3\ \mu} +\xi m_W \varphi_2)^2
-\nonumber\\ -{1\over 2\xi}
(\partial^\mu A_\mu^3 + \xi m_Z \cos\theta_W\varphi_3)^2-
{1\over 2\xi}
(\partial^\mu A_\mu^0 - \xi m_Z \sin\theta_W\varphi_3)^2
\end{eqnarray}
which generalizes the linear $R_\xi$-gauge and is more suitable
for calculations in a background electromagnetic field
\cite{NL-gauge}. $A_\mu^a$ and $A_\mu^0$ denote here the $SU(2)$
and $U(1)$ gauge fields respectively whereas $\varphi_a$
designates the Goldstone bosons. The Lagrangian (\ref{L-W})
brings the equation of motion
\begin{equation}
\label{W-equation}
\Bigl[(\partial_\nu\pm ieA_\nu)^2 + m_W^2 \Bigr]W^{\pm\mu}+
(1/\xi-1)(\partial^\mu \pm ieA^\mu)(\partial^\nu W^{\pm}_\nu \pm
ieA^\nu W^{\pm}_\nu) \pm 2ieF^{\mu\nu} W^{\pm}_\nu = 0
\end{equation}
describing charged massive particles with spin $0$ and $1$. The
longitudinal mode of the vector field $W^{\pm}$
\begin{equation}
\label{B-field}
B^{\pm}(x)= -{1\over \xi m_W}(\partial^\mu W^{\pm}_\mu \pm
ieA^\mu W^{\pm}_\mu)
\end{equation}
corresponds to the unphysical scalar with mass $\xi m_W$ and it
obeys the Klein-Gordon equation
\begin{equation}
\label{B-equation}
\Bigl[(\partial_\nu\pm ieA_\nu)^2 +\xi m_W^2 \Bigr] B^\pm (x) = 0.
\end{equation}
The wave functions of $W$-bosons come from the transversal mode
being associated with vector quantum states.
\begin{equation}
\label{V-field}
V^{\pm}_\mu (x)= W^{\pm}_\mu (x)+ {1\over \xi m_W^2}(\partial_\mu
\pm ieA_\mu)(\partial^\nu W^{\pm}_\nu \pm ieA^\nu W^{\pm}_\nu)
\end{equation}
The equation for the transversal mode $V^{\pm}_\mu$ is derived
from eq.(\ref{W-equation}) and it can be written in the
following way.
\begin{equation}
\label{V-equation}
\Bigl[(\partial_\nu\pm ieA_\nu)^2 + m_W^2 \Bigr] V^{\pm\mu}(x)
\pm 2ieF^{\mu\nu} V^{\pm}_\nu (x) = 0
\end{equation}
Besides, due to eq.(\ref{B-equation}) the field $V^{\pm}_\mu$
satisfies additional condition
\begin{equation}
(\partial^\mu \pm ieA^\mu) V^{\pm}_\mu (x) = 0
\end{equation}
which means that only $3$ components of the field $V^{\pm}_\mu$
are independent (exactly as it should be for spin-1 particles).
Thus, we have obtained a Lorentz-covariant decomposition of the
field $W^{\pm}_\mu$ into the scalar and vector parts:
\begin{equation}
\label{W-field}
W^{\pm}_\mu (x)= V^{\pm}_\mu (x)+ {1\over m_W}
(\partial_\mu \pm ieA_\mu) B^{\pm}(x)
\end{equation}
Notice that all physically sensible information is contained in
the vector field $V^{\pm}_\mu$, whereas the scalar field
$B^{\pm}$ plays an auxiliary role and its contribution to the
S-matrix is canceled by that of the Goldstone bosons $\varphi^{\pm}=
{1\over \sqrt{2}}(\varphi_2 \pm i\varphi_1)$.

To carry out the scheme of canonical quantization, one needs an
explicit form of the $W$-boson wave functions in an intense
electromagnetic background. This requires obtaining exact
solutions of eq.(\ref{V-equation}) for a given configuration of
the external field. If we restrict ourselves to the case of a
plane-wave electromagnetic field with potential $A_\mu(\varphi)$
then eq.(\ref{V-equation}) can be easily integrated and we get:
\begin{eqnarray}
\label{W-function}
W^{\pm}_{\mu}(x,{\bf p})=\exp\biggl[-ipx-i\int\limits_0^{kx}
\biggl(\pm\frac{epA}{pk}-\frac{e^2 A^2}{2pk}\biggr)d\varphi\biggr]
\cdot\nonumber\\
\biggl[g_{\mu\nu}\pm\frac{e}{pk}\biggl(k_\mu A_\nu (kx)-
k_\nu A_\mu (kx) \biggr) - \frac{e^2}{2(pk)^2} A^2(kx)
k_\mu k_\nu \biggr]\frac{v^{\pm\nu}(p,\sigma)}{\sqrt{2p_0 V}},
\end{eqnarray}
where $v^{\pm\nu}(p,\sigma)$ is a complex polarization vector
which is subject to the following constraints:
\begin{equation}
\label{W-polarization}
v^{\pm}_{\mu} p^\mu=0, \qquad v^{\pm *}_\mu v^{\pm\mu}=-1, \qquad
\sum_{\sigma} v^{\pm}_{\mu}(p,\sigma) v^{\pm *}_{\nu}(p,\sigma)=
- g_{\mu\nu} + p_\mu p_{\nu}/m_W^2.
\end{equation}
Now the task is to represent the field $W^{\pm}_{\mu}$ as the sum
of one-particle quantum excitations. Using expression
(\ref{W-field}) we expand $W^{\pm}_\mu$ into series of classical
solutions (\ref{W-function}), (\ref{scalar-wf})
\begin{equation}
\label{quant-W+}
W^+_\mu (x)=\sum_{{\bf p}\sigma}\Bigl[a_{{\bf p}\sigma}
W^+_\mu (x,{\bf p})+ b_{{\bf p}\sigma}^{\dagger}
W^{- *}_\mu (x,{\bf p})\Bigr] +
\frac{(\partial_\mu + ie A_\mu)}{m_W}
\sum_{\bf p} \Bigl[a_{{\bf p} 0} \phi^+(x,{\bf p})+
b_{{\bf p} 0}^{\dagger} \phi^{- *} (x,{\bf p})\Bigr],
\end{equation}
\begin{equation}
\label{quant-W-}
W^-_\mu (x)=\sum_{{\bf p}\sigma}\Bigl[a_{{\bf p}\sigma}^{\dagger}
W^{+ *}_\mu (x,{\bf p}) + b_{{\bf p}\sigma}
W^{-}_\mu (x,{\bf p})\Bigr] +
\frac{(\partial_\mu - ie A_\mu)}{m_W}
\sum_{\bf p}\Bigl[a_{{\bf p} 0}^{\dagger}\phi^{+ *}(x,{\bf p})+
b_{{\bf p} 0}\phi^{-} (x,{\bf p})\Bigr],
\end{equation}
which incorporate the Fock operators of two kinds. The observables
$a_{{\bf p}\sigma}, b_{{\bf p}\sigma}$ correspond to the vector
quantum states of $W$-bosons and satisfy the usual commutation
relations (with a polarization subscript $\sigma =1,2,3$.)
\begin{eqnarray}
\label{commutators-3}
\bigl[ a_{{\bf p}\sigma};a_{{\bf p'}\sigma'}^{\dagger}\bigr]=
\delta_{\bf p p'}\delta_{\sigma\sigma'} \qquad
\bigl[ b_{{\bf p}\sigma};b_{{\bf p'}\sigma'}^{\dagger}\bigr]=
\delta_{\bf p p'}\delta_{\sigma\sigma'} \nonumber\\
\bigl[ a_{{\bf p}\sigma};a_{{\bf p'}\sigma'} \bigr]=
\bigl[ b_{{\bf p}\sigma};b_{{\bf p'}\sigma'} \bigr]=
\bigl[ a_{{\bf p}\sigma};b_{{\bf p'}\sigma'} \bigr]=
\bigl[ a_{{\bf p}\sigma};b_{{\bf p'}\sigma'}^{\dagger}\bigr]=0
\end{eqnarray}
At the same time the operators $a_{{\bf p} 0}, b_{{\bf p} 0}$
connected with the scalar part (\ref{B-field}) must obey different
rules of quantization:
\begin{eqnarray}
\label{commutators-4}
\bigl[ a_{{\bf p} 0};a_{{\bf p'}0}^{\dagger}\bigr]=
- \delta_{\bf p p'} \qquad
\bigl[ b_{{\bf p} 0};b_{{\bf p'} 0}^{\dagger}\bigr]=
- \delta_{\bf p p'} \nonumber\\
\bigl[ a_{{\bf p} 0};a_{{\bf p'} 0} \bigr]=
\bigl[ b_{{\bf p} 0};b_{{\bf p'} 0} \bigr]=
\bigl[ a_{{\bf p} 0};b_{{\bf p'} 0} \bigr]=
\bigl[ a_{{\bf p} 0};b_{{\bf p'} 0}^{\dagger}\bigr]=0
\end{eqnarray}
The odd minus in eq.(\ref{commutators-4}) is a price that must
be paid for the Lorentz-covariance. It immediately causes the
appearance of indefinite metrics in Hilbert space ${\cal H}$
which is typical for Gupta-Bleuler quantization. However, since
longitudinal mode $B^{\pm}$ (\ref{B-field}) makes no physical
sense, one can eliminate it by considering the physical subspace
${\cal H}_{phys} \subset {\cal H}$ without these scalar
excitations. Similar arguments can be found in many textbooks on
quantum field theory \cite{Itzykson-Zuber}. Loop calculations in
a background electromagnetic field involve Feynman propagator
being defined by the formula:
\begin{equation}
\label{propagator-1W}
D_{\mu\nu}(x,x')=-i<0\vert T W^+_\mu (x) W^-_\nu (x')\vert 0>,
\end{equation}
and which is a Green function of the equation (\ref{W-equation})
\begin{equation}
\label{Green-W}
\Bigl[\bigl((\partial_{\alpha} + ieA_{\alpha})^2 + m_W^2 \bigr)
g^{\mu\nu}+(1/\xi-1)(\partial^\mu+ieA^\mu)(\partial^\nu+ieA^\nu)
+2ieF^{\mu\nu}\Bigr] D_{\nu\lambda}(x,x')=\delta^{\mu}_{\lambda}
\delta^4(x-x')
\end{equation}
Inserting expansions (\ref{quant-W+}) and (\ref{quant-W-}) into
eq.(\ref{propagator-1W}) and employing the commutation rules
(\ref{commutators-3}), (\ref{commutators-4}), one can obtain an
explicit covariant form of the $W$-boson propagator:
\begin{equation}
\label{propagator-2W}
D_{\mu\nu} (x,x')=\int\frac{d^4p}{(2\pi)^4}
\frac{W_{\mu\alpha}(x,p) W^*_{\beta\nu}(x',p)}{m^2_W-p^2-i0}
\biggl[g^{\alpha\beta}+(\xi-1)\frac{p^{\alpha}p^{\beta}}
{p^2-\xi m_W^2 +i0}\biggr].
\end{equation}
This representation is based on the eigen functions of the
differential operator corresponding to eq.(\ref{V-equation})
\begin{eqnarray}
\label{eigen-f}
W_{\mu\alpha}(x,p)=\exp\biggl[-ipx-i\int\limits_0^{kx}
\biggl(\frac{epA}{pk}-\frac{e^2 A^2}{2pk}\biggr)d\varphi\biggr]
\cdot\nonumber\\ \biggl\{g_{\mu\alpha}+\frac{e}{pk}\biggl(k_\mu
A_\alpha (kx)- k_\alpha A_\mu (kx) \biggr) - \frac{e^2}{2(pk)^2}
A^2(kx) k_\mu k_\alpha \biggr\}
\end{eqnarray}
Let us note that classical solution (\ref{W-function}) is simply a
particular form of the eigen function (\ref{eigen-f}) taken on the
mass shell and contracted through the last index with the ordinary
polarization vector. For an arbitrary value of gauge fixing
parameter $\xi$ a proper time representation of the propagator
(\ref{propagator-1W}) is rather awkward. However, practically one
can use a diagonal gauge ($\xi=1$) in which expression
(\ref{propagator-2W}) can be rewritten in the most elegant way:
\begin{eqnarray}
\label{propagator-3W}
D_{\mu\nu}(x,x')=\frac{\Phi(x,x')}{(4\pi)^2}\int\limits_0^\infty
\frac{ds}{s^2}\exp\biggl[-ism^2-i\frac{(x-x')^2}{4s}
+\frac{ie^2s}{(kx-kx')}\int\limits_{kx'}^{kx}A_\mu(\varphi)
A^\mu(\varphi)d\varphi-\nonumber\\
-\frac{ie^2s}{(kx-kx')^2}\int\limits_{kx'}^{kx}
d\varphi_1\int\limits_{kx'}
^{kx}d\varphi_2A_\mu(\varphi_1)A^\mu(\varphi_2)\biggr]
\biggr\{g_{\mu\nu} + \frac{2es}{(kx-kx')}
\Bigl[k_\mu \Bigl( A_\nu(kx)-A_\nu(kx')\Bigr)-
\nonumber\\
- k_\nu \Bigl( A_\mu(kx)- A_\mu(kx') \Bigr)\Bigr]
-\frac{2e^2 s^2}{(kx-kx')^2}\Bigl( A_\lambda (x)-
A_\lambda (x') \Bigr)^2 k_\mu k_\nu\biggr\}
\end{eqnarray}

\section{General Features of Quantum Processes in Strong
Electromagnetic Fields}

Many distinctive properties of particle physics with intense
electromagnetic background can be traced in the reaction of
$W$-boson and neutrino production via the decay of a massive
charged lepton $\ell$. It is well known that the process
$\ell^-\rightarrow W^-\nu_\ell$ should be forbidden in vacuum if
the lepton mass is small in comparison with the ones of final
particles. However, in the presence of electromagnetic fields the
restriction $m_\ell > m_W + m_\nu$ can be removed and there
appears a possibility to observe an effect analogous to quantum
transmission of the potential barrier through tunneling. The
matrix element of the reaction $\ell^-\rightarrow W^-\nu_\ell$ can
be calculated employing the wave functions of the charged lepton
$\ell$ (\ref{spinor-wf}) and the one of W-boson
(\ref{W-function}):
\begin{equation}
\label{S-matrix}
S_{fi}=\frac{ig}{2\sqrt{2}}\int d^4 x W^{-*}_\mu (x,p')
\overline{\nu_\ell}(x,p'')\gamma^\mu (1+\gamma^5)
\Psi^{-}_\ell (x,p)
\end{equation}
In order to obtain the probability of the decay $\ell^-\rightarrow
W^-\nu_\ell$ per unit of time one must integrate $\vert
S_{fi}\vert^2$ over the phase volume of final particles and sum up
their polarizations. In the case of a constant crossed field
(${\cal F}={\cal G}=0$) the final result can be written as
\cite{Kurilin-88}:
\begin{eqnarray}
\label{prob-W}
\frac{d P}{du}(\ell^-\rightarrow W^-\nu_\ell)=-\frac{G_F m_W^4}
{4\pi^2\sqrt{2} p_0} \biggl\{ \biggl[1-\frac{(m_\ell^2 + m_\nu^2)}
{2 m_W^2} - \frac{(m_\ell^2 - m_\nu^2)^2}{2m_W^4}\biggr]
\Phi_1(z)+\nonumber\\
+\frac{2 m_e^2(\chi^2 u)^{1/3}}{m_W^2 (1-u)^{4/3}}
\biggl[1+u^2 + (1-u)^2\frac{(m_\ell^2+m_\nu^2)}{2m_W^2}\biggr]
\Phi'(z) \biggr\}
\end{eqnarray}
This formula represents the differential probability of the
process with respect to the variable
\begin{equation}
\label{u}
u=\frac{(p''_\mu k^\mu)}{(p_\lambda k^\lambda)}=
\frac{(p''_\mu F^{\mu\alpha} F_{\alpha\beta} p^\beta)}
{(p_\mu F^{\mu\alpha} F_{\alpha\beta} p^\beta)},
\end{equation}
where $p''_\mu, p'_\mu$ are momenta of neutrino and $W$-boson,
respectively, while $p_\mu$ stands for the one of initial lepton.
Note that in a crossed electromagnetic field $F_{\mu\lambda}$ the
wave vector $k_\mu$ must be substituted by the following
expression:
\begin{equation}
\label{k-vector}
k^\mu =\frac{e^2(pk)}{m_e^6 \chi^2} F^{\mu\alpha}F_{\alpha\beta}
p^\beta \sim \biggl( p_0 {\bf E}^2-{\bf p}({\bf E}\times{\bf B});
\ \ {\bf E}({\bf E p})+{\bf B}({\bf B p})-{\bf p}({\bf B B})+
p_0({\bf E}\times {\bf B}) \biggr).
\end{equation}
This interrelation meaning that relativistic particles perceive
the electromagnetic background as an incident plane-wave have been
noticed long ago by Williams and Weizs\"acker \cite{EPA} who
proposed the famous equivalent-photon approximation (see also
\cite{EPA-new}). In the case of a constant homogeneous
electromagnetic field being considered the kinematics of the
reaction $\ell^-\rightarrow W^-\nu_\ell$ resembles the one with
real photons and is determined by the following constraint of
4-momentum conservation:
\begin{equation}
\label{kinematics}
p_\mu + \frac{e^2 M^2}{2 m_e^6 \chi^2} F_{\mu\alpha} F^{\alpha\beta}
p_\beta= p'_\mu + p''_\mu,
\end{equation}
where $M^2$ represents the energy deficit obstructing the decay
$\ell^-\rightarrow W^-\nu_\ell$ in a vacuum without fields.
\begin{equation}
\label{deficit}
M^2=(p'+p'')^2-p^2=m_W^2+m_\nu^2-m_\ell^2+2(p'p'')\geq
(m_W + m_\nu)^2 - m_\ell^2
\end{equation}
Now we see that for light lepton ($m_\ell < m_W + m_\nu$) $M^2$
is always positive and the reaction could proceed only due to
the energy-momentum borrowed from the background field.
\begin{equation}
\label{field-energy}
q^\mu=\frac{M^2}{2(pFFp)} F^{\mu\alpha} F_{\alpha\beta} p^\beta
\qquad\qquad
q^2=\frac{M^4}{4 m_e^2}\biggl[ \biggl({a\over\chi^2}\biggr)^2-
{2b\over\chi^2}\biggr]\approx 0
\end{equation}
If we suppose the lepton $\ell$ to be a hypothetical heavy
particle of the fourth generation ($m_\ell > m_W + m_\nu$) then
for some values of neutrino and $W$-boson momenta $M^2$ can
become negative. This reflects another possibility when the
background field takes some amount of energy from the initial
lepton and absorbs it. Thus electromagnetic environment reveals
itself as an additional virtual particle with momentum $q^\mu$
(\ref{field-energy}) that induces the decay $\ell^-\rightarrow
W^-\nu_\ell$ regardless whether it is forbidden or not when the
external field is absent.

All said above is described by the formula (\ref{prob-W}) which
involves special functions
\begin{equation}
\Phi(z)=\int\limits_0^\infty dt\cos(zt+t^3/3),\qquad
\Phi_1(z)=\int\limits_z^\infty dt \Phi(t),\qquad  \Phi'(z)=
\frac{d\Phi(z)}{dz}
\end{equation}
which are known in mathematics as the functions of Airy (see the
Appendix). The argument $z$ is proportional to the quantity $M^2$
(\ref{deficit}) and in terms of spectral variable $u$ (\ref{u}) it
can be written as follows:
\begin{equation}
\label{z}
z=\frac{m_W^2 u + m_\nu^2 (1-u) - m_\ell^2 u(1-u)}
{m_e^2 [\chi u^2 (1-u)]^{2/3}}
\end{equation}
Note that due to the constraint (\ref{kinematics}) the momentum
projection on the wave vector (\ref{k-vector}) is conserved
\begin{equation}
\label{projection}
p_\mu
F^{\mu\alpha} F_{\alpha\beta} p^\beta= (p'_\mu + p''_\mu)
F^{\mu\alpha} F_{\alpha\beta} p^\beta
\end{equation}
that fixes the interval of variations for $u$: $0\le u \le 1$. Now
if the reaction $\ell^-\rightarrow W^-\nu_\ell$ is forbidden in
vacuum, then $z>0$ for all $u\in [0;1]$. In the opposite case:
$m_\ell > m_W + m_\nu$ there is a domain of negative values
\begin{eqnarray}
z<0 \qquad for\qquad  u_1 < u < u_2, \nonumber\\
u_{2,1}=\frac{1}{2 m_\ell^2}\biggl[m_\ell^2+m_\nu^2-m_W^2
\pm
\sqrt{[m_\ell^2-(m_W+m_\nu)^2][m_\ell^2-(m_W-m_\nu)^2]}\biggr],
\end{eqnarray}
which exactly coincide with the phase volume of the process
at hand provided that external field is omitted.

The sign and magnitude of the argument $z$ exert primary control
over the probability of any quantum process in the background
electromagnetic field. This is caused by a particular behaviour
of the Airy functions that is examined carefully in the Appendix.
Let us only mention that for large positive values of $z$
these functions decrease exponentially as $\exp(-{2\over 3}
z^{3/2})$. So the probabilities of reactions being forbidden
under normal conditions must contain a factor of exponential
suppression. This exponent can be treated as the well-known
semiclassical probability of quantum transmission below the
potential energy barrier. The index of the power increasing
proportionally with $M^2$ (\ref{deficit}) characterizes how much
energy is borrowed from the background field. The aforesaid
arguments can be confirmed by the asymptotic estimate of the
total probability of the decay $\ell^-\rightarrow W^-\nu_\ell$
in the domain of weak electromagnetic fields. Integrating
eq.(\ref{prob-W}) through the use of the saddle-point method one
obtains.
\begin{equation}
\label{prob-W1}
P(\ell^-\rightarrow W^-\nu_\ell)=\frac{2G_F m_W m_e^3 \chi}
{9\pi\sqrt{6} p_0} \exp\biggl(-\frac{\sqrt{3}m_W^3}
{\chi m_e^3}\biggr), \quad for
\quad \chi\ll \biggl({m_W\over m_e}\biggr)^3
\end{equation}
Here it is implied that masses of particles involved in the
reaction satisfy real phenomenological conditions ($m_\nu\ll
m_\ell\ll m_W$). We see that all leptons being known up to now can
produce $W$-bosons only if the field strength $E$ and lepton
energy $p_0$ take the values that are comparable with the
estimate:
\begin{equation}
\label{estimate-W}
\chi=\biggl({p_0\over m_e}\biggr)\biggl({E\over E_{cr}}\biggr)
\sim \sqrt{3}\biggl({m_W\over m_e}\biggr)^3
\approx 6.7\cdot 10^{15}
\end{equation}
In other words this exotic process could occur perhaps only in the
Early Universe when the energies $p_0\sim 10^{12} GeV$  were
accessible. This ultra-relativistic domain is described by the
formula which is derived from eq.(\ref{prob-W}) by means of the
expansions (\ref{series-1}), (\ref{series-2}) (see the Appendix).
\begin{equation}
\label{prob-W2}
P(\ell^-\rightarrow W^-\nu_\ell)=\frac{3G_F m_W m_e^3 \chi}
{2\pi\sqrt{6} p_0}, \quad for
\quad \chi\gg \biggl({m_W\over m_e}\biggr)^3
\end{equation}
The above results can be summed up as the following statement.
If the reaction proceeding by the background electromagnetic
field  is forbidden in vacuum, then for weak fields its
probability is exponentially suppressed by the factor which has
the lacking energy borrowed from the background field as the
power index. This suppression becomes unimportant only in the
domain of extremely strong fields and ultra-relativistic energies
that practically are inaccessible in experiments, at least now.

The situation changes dramatically for the processes that could
take place when the background field is absent. For example, if we
consider the decay of a hypothetical heavy lepton of the fourth
generation ($m_\ell > m_W + m_\nu$), then from eq.(\ref{prob-W})
it follows.
\begin{eqnarray}
\label{prob-W3}
P(\ell^-\rightarrow W^-\nu_\ell)=\frac{G_F\sqrt{2}}
{16\pi p_0} \sqrt{
\biggl[1-\biggl({m_W+m_\nu\over m_\ell}\biggr)^2\biggr]
\biggl[1-\biggl({m_W-m_\nu\over m_\ell}\biggr)^2\biggr]}
\nonumber\times\\
\Bigl[ (m_\ell^2-m_\nu^2)^2 +m_W^2 (m_\ell^2+m_\nu^2)
-2m_W^4 \Bigr] + O(\chi^2)
\end{eqnarray}
This expression exactly reproduces the probability of the reaction
$\ell^-\rightarrow W^-\nu_\ell$ which can be calculated through
the standard quantum field theory technique. Note, that it arises
from the formal substitution $\Phi_1(z) \rightarrow
\pi\theta(-z)$\  ($\theta(x)$ is the Heaviside step function)
giving the first term in the asymptotic expansion of
eq.(\ref{prob-W}) when $\chi\rightarrow 0$. Thus we see that the
probability of the permitted in vacuum process acquires in the
background field corrections proportional to $\chi^2$ (the omitted
term $O(\chi^2)$ in eq.(\ref{prob-W3})).  However these
corrections which become essential  only in the domain
(\ref{estimate-W}) can be neglected if $\chi\ll (m_W/m_e)^3$.

\section{New Processes in Background Electromagnetic Fields}

Now I would like to discuss the possibilities for searching new
physics in background electromagnetic fields. We have already
noticed that the relativistic particle penetrating through the
electromagnetic environment feels the background field as a beam
of real photons which can supply it with substantial amount of
energy. If the field strength is sufficiently high then new
channels of reactions can be opened and the particle decays into
more heavy species. Let us consider such a reaction when a light
charged particle $A^\pm$ produces two others, one of which is
charged too ($B^\pm$) while the other is neutral ($C^0$). We
assume that the process $A^\pm \rightarrow B^\pm C^0$ is forbidden
in a vacuum ($ m_A < m_B + m_C$) and wish to estimate the
field-energy domains where it could occur. As we have explained in
the previous section, in general the probability of the decay
$A^\pm \rightarrow B^\pm C^0$ having a form similar to
eq.(\ref{prob-W}) can be represented through the Airy functions
which depend on the following argument:
\begin{equation}
\label{Z-charged}
Z_\pm=\frac{m_B^2 u + m_C^2 (1-u) - m_A^2 u(1-u)}
{m_e^2 [\chi u^2 (1-u)]^{2/3}}
\end{equation}
For weak electromagnetic fields that can be handled in
experiments the above probability must contain the exponential
factor which is connected with the minimal positive value of the
argument $Z_\pm$ (\ref{Z-charged}):
\begin{equation}
P(A^\pm \rightarrow B^\pm C^0) \sim \exp(-{2\over 3}
Z_{\pm\  min}^{3/2})
\equiv \exp(-\gamma)
\end{equation}
Using eq.(\ref{Z-charged}) one can explicitly calculate the
power index $\gamma$ for the arbitrary relationship among
the masses of all particles involved in the reaction:
\begin{equation}
\label{gamma}
\gamma=\frac{\sqrt{6}}{16\chi}
\biggl\{ 128\lambda_1^2 \lambda_2 + 80\lambda_1
(\lambda_1+\lambda_2-\lambda_0)^2 +
\frac{(\lambda_1+\lambda_2-\lambda_0)^4}{\lambda_2}
\biggl[ \biggl(1+\frac{32\lambda_1 \lambda_2}
{(\lambda_1+\lambda_2-\lambda_0)^2}\biggr)^{3/2}
-1 \biggr] \biggr\}^{1/2},
\end{equation}
where
\begin{equation}
\label{lambda-012}
\lambda_0=\biggl({m_A\over m_e}\biggl)^2, \qquad
\lambda_1=\biggl({m_B\over m_e}\biggl)^2, \qquad
\lambda_2=\biggl({m_C\over m_e}\biggl)^2
\end{equation}
This quantity governing the range of exponential suppression makes
a crucial impact on the probability and fixes a certain threshold
of the reaction induced by the electromagnetic background. In the
domain of the very small values of parameter $\chi$ the
probability of the decay $A^\pm \rightarrow B^\pm C^0$ becomes
negligible on the account of extremely large magnitude of the
power index ($\gamma\gg 1$). So the reaction could proceed with a
feasible rate only if $\gamma\sim 1$. I will refer to this
condition as to the threshold of the reaction designating the
corresponding value of $\chi$ by a zero subscript ($\chi_0$).

Let us analyze the consequences of the general expression
(\ref{gamma}). Suppose that neutral particle being produced in
the decay $A^\pm \rightarrow B^\pm C^0$ is very light in
comparison with the others. Then the power index $\gamma$ can be
approximated as follows.
\begin{equation}
\label{gamma-1}
\gamma=\frac{\sqrt{3}}{\chi}\biggl({m_B\over m_e}\biggr)^3
\biggl( 1- \frac{m_A^2}{m_B^2}\biggr),
\qquad for \quad m_C \ll m_A, m_B.
\end{equation}
A particular case of this formula have been employed in the
previous section for the process of $W$-boson and neutrino
production. The threshold of this reaction
$\ell^-\rightarrow W^-\nu_\ell$ coinciding with the estimate
(\ref{estimate-W}) can be calculated without resort to the
explicit form of the probability (\ref{prob-W1}). The only thing
that we need is the power index (\ref{gamma-1}) being connected
with the threshold value of $\chi$ by a simple equation
$\gamma=1$. This reasoning can be extended to some other
reactions being of practical interest. Substituting the values
of electron, muon and pion masses in eq.(\ref{gamma-1}) we obtain
characteristic domains of the following processes induced by the
external electromagnetic background.
\begin{eqnarray}
\label{leptons}
e^-\rightarrow\mu^-\nu_e\tilde\nu_\mu,\
e^+\rightarrow\mu^+\tilde\nu_e \nu_\mu,
\qquad \chi_0=1.53\cdot 10^7 \nonumber\\
e^-\rightarrow\pi^-\nu_e,\
e^+\rightarrow\pi^+\tilde\nu_e,
\qquad \chi_0=3.53\cdot 10^7 \nonumber\\
\mu^-\rightarrow\pi^-\nu_\mu,\
\mu^+\rightarrow\pi^+\tilde\nu_\mu,
\qquad \chi_0=1.51\cdot 10^7
\end{eqnarray}
On the other hand, if we assume that the mass of the charged
particle emerged in the decay $A^\pm \rightarrow B^\pm
C^0$ is the smallest parameter, then eq.(\ref{gamma}) can be
reduced to
\begin{equation}
\label{gamma-2}
\gamma=\frac{\sqrt{3}m_B}{\chi m_e}\biggl({m_C\over m_e}\biggr)^2
\biggl( 1- \frac{m_A^2}{m_C^2}\biggr),
\qquad for \quad m_B \ll m_A, m_C.
\end{equation}
Now it is evident that the factor of exponential suppression grows
up from the non-zero mass of the charged particle being produced,
because the power index $\gamma$ (\ref{gamma-2}) vanishes in the
limit $m_B \rightarrow 0$. By virtue of the fact that there are no
charged particles with masses less than the one of electron, it is
reasonable to look for the reactions with electrons and positrons
in final states. For example, the processes of the inverse beta
decay being forbidden in vacuum have the threshold that lies far
below the ones of the lepton reactions (\ref{leptons}). (This is
also caused by a small mass difference between proton and
neutron.)
\begin{equation}
\label{inverse-beta}
p^+\rightarrow n e^+\nu_e,\
\tilde p^-\rightarrow \tilde n e^-\tilde\nu_e ,
\qquad \chi_0=1.61\cdot 10^4
\end{equation}

Another interesting possibility arises in connection with
the process of neutral particle emission in the intense
electromagnetic field $A^\pm \rightarrow A^\pm C^0$.
The power index $\gamma$ that determines the rate of this
reaction can be deduced from eq.(\ref{gamma}) by a formal
replacement $m_B=m_A$.
\begin{equation}
\label{gamma-3}
\gamma=\frac{\sqrt{6}}{16\chi}\biggl({m_C\over m_e}\biggr)^3
\biggl[\biggl(1+32 {m_A^2\over m_C^2}\biggl)^{3/2}-1+
80\biggl({m_A\over m_C}\biggl)^2
+128\biggl({m_A\over m_C}\biggl)^4 \biggr]^{1/2}.
\end{equation}

Now provided that the neutral particle is much heavier than its
fore-runner we can rewrite eq.(\ref{gamma-3}) in the form.
\begin{equation}
\label{gamma-31}
\gamma=\frac{\sqrt{3}m_A}{\chi m_e}\biggl({m_C\over m_e}\biggr)^2,
\qquad for \quad m_A \ll m_C.
\end{equation}
Here it is relevant to quote the reactions of pion radiation as an
example governed by the last formula.
\begin{equation}
\label{e-pion}
e^-\rightarrow e^-\pi^0,\
e^+\rightarrow e^+\pi^0,
\qquad \chi_0=1.21\cdot 10^5.
\end{equation}

In the opposite case corresponding to the
relatively small mass of the emitted particle the factor of
exponential suppression appears with the following index.
\begin{equation}
\label{gamma-32}
\gamma=\frac{\sqrt{3}m_C}{\chi m_e}\biggl({m_A\over m_e}\biggr)^2,
\qquad for \quad m_A \gg m_C.
\end{equation}
So the process of pion radiation from protons could occur only
if the field strength and proton energy exceed significantly the
ones of the reactions (\ref{e-pion}).
\begin{equation}
\label{p-pion}
p^+\rightarrow p^+\pi^0,\
\tilde p^-\rightarrow \tilde p^-\pi^0,
\qquad \chi_0=8.90\cdot 10^8.
\end{equation}

Next we intend to discuss the possibilities of charged particles
production via the field induced decay $A^0 \rightarrow B^+ C^-$.
The argument of the Airy functions through which the probability
of this reaction can be expressed has a form that slightly differs
from eq.(\ref{Z-charged}).
\begin{equation}
\label{Z-neutral}
Z_0=\frac{m_B^2 u + m_C^2 (1-u) - m_A^2 u(1-u)}
{m_e^2 [\chi u (1-u)]^{2/3}}
\end{equation}
Following the reasoning given above it is straightforward to
calculate the exponential factor which governs the probability
rate in the domain of weak electromagnetic fields.
\begin{equation}
\label{neutral-decay}
P(A^0 \rightarrow B^+ C^-) \sim \exp(-{2\over 3} Z_{0\ min}^{3/2})
\equiv\\exp(-\gamma)
\end{equation}
At this stage, in order to obtain precise estimates of the
threshold of the process $A^0 \rightarrow B^+ C^-$ we need
additional assumptions concerning with the masses of particles
being implicated. If the initial neutral particle $A^0$ has a
very small mass as compared to the charged ones
($m_A \ll m_B, m_C$) then the power index of
eq.(\ref{neutral-decay}) can be written as follows:
\begin{equation}
\label{gamma-4}
\gamma=\frac{\sqrt{2}}{6\chi}
\biggl[ (\lambda_1^2+14 \lambda_1\lambda_2+\lambda_2^2 )^{3/2}
-\lambda_1^3 + 33\lambda_1^2\lambda_2
+ 33\lambda_1\lambda_2^2 -\lambda_2^3 \biggr]^{1/2}
\end{equation}
The last equation describes, for example, the reactions being
caused by neutrinos moving throughout the intense electromagnetic
field. This is especially the case when one considers the process
which is cross symmetric to the muon decay.
\begin{equation}
\label{mu-neutrino}
\nu_\mu\rightarrow e^+\mu^-\nu_e,\
\tilde\nu_\mu\rightarrow e^-\mu^+\tilde\nu_e
\qquad \chi_0=7.41\cdot 10^4
\end{equation}
Confronting the conditions appropriate to the above reaction with
the threshold of another cross symmetric channel adduced in
eq.(\ref{leptons}) we see that reactions initiated by neutrinos
require much less values of energy and background field strength
than the ones with electrons. From this viewpoint  it is more
promising to look for new physics in external electromagnetic
fields through the investigations of the processes induced by
neutral particles. This conclusion acquires further confirmation
from the asymptotic estimate of the power index
(\ref{neutral-decay}) in the case when one of charged particles
produced is extremely light.
\begin{equation}
\label{gamma-5}
\gamma=\frac{\sqrt{3}m_C (m_B^2-m_A^2)}{\chi m_e^3},
\qquad for \qquad m_C \ll m_A, m_B.
\end{equation}
Related conditions are characteristic for the following pion
decays, which have the lowest thresholds among the all
afore-mentioned reactions.
\begin{equation}
\label{pi-0}
\pi^0\rightarrow \pi^+e^-\tilde\nu_e,\
\pi^0\rightarrow \pi^-e^+\nu_e
\qquad \chi_0=8.38\cdot 10^3 ,
\end{equation}

It should also be noted that the power index (\ref{gamma-5})
exactly coincides with the estimate (\ref{gamma-2}) if the
interchange $m_B \leftrightarrow m_C$ is made. This fact explains
once more the origin of the exponential suppression as a
semiclassical probability of the field-induced quantum
transmission of the electron from the Dirac sea to the upper
continuum. The height of the potential barrier corresponding to
the results (\ref{gamma-2}), (\ref{gamma-5}) can be easily
calculated if one reproduces the reasoning which is relevant to
the famous Klein paradox in quantum mechanics \cite{Klein}.

Finally, I would like to make some remarks about the processes
of particle-antiparticle pairs production in the intense
electromagnetic fields. The probability of the forbidden in a
vacuum decay $A^0 \rightarrow B^+B^-$ is suppressed in the
electromagnetic background by the factor which can be obtained
from eq.(\ref{Z-neutral}), (\ref{neutral-decay}) under the
assumption $m_C=m_B$ :
\begin{equation}
P(A^0 \rightarrow B^+B^-) \sim
\exp\biggl[-\frac{8 m_B^3}{3\chi m_e^3}\biggl(1-\frac{m_A^2}
{4 m_B^2} \biggr)^{3/2}\biggr]
\end{equation}
This formula can be applied for the threshold evaluations
related to the following reactions provoked by the external field.
\begin{equation}
\label{e-pairs}
\gamma\rightarrow e^+e^-,\
\nu_\mu\rightarrow \nu_\mu e^+e^-
\qquad \chi_0=2.67
\end{equation}
\begin{equation}
\label{mu-parirs}
\gamma\rightarrow \mu^+\mu^-, \
\nu_\mu\rightarrow \nu_\mu \mu^+\mu^-
\qquad \chi_0=2.36\cdot 10^7.
\end{equation}
We see that electron-positron pairs can be easily produced at the
energies $p_0 > 25 \ GeV$ if the background field strength is of
the order $E\sim 10^{10} V/m$. Similar conditions have been
created in CERN experiments \cite{CERN 94-05} for the photons
incident on the crystal of Germanium with a small angle to the
axial direction. Neutrino fluxes of the like energies could give
rise to the same effect caused by the coherent interactions with a
regular arrangement of atoms in the single crystal. However due to
the very small values of the neutrino couplings the process
$\nu_\mu\rightarrow \nu_\mu e^+e^-$ has an additional suppression
by the factor $(m_e/m_W)^4 \sim 10^{-21}$ \cite{Borisov-93} which
impedes its immediate experimental observation.

\begin{figure}[t]\label{chi-fig}
\setlength{\unitlength}{1cm}
\begin{center}
\epsfxsize=15.cm \epsffile{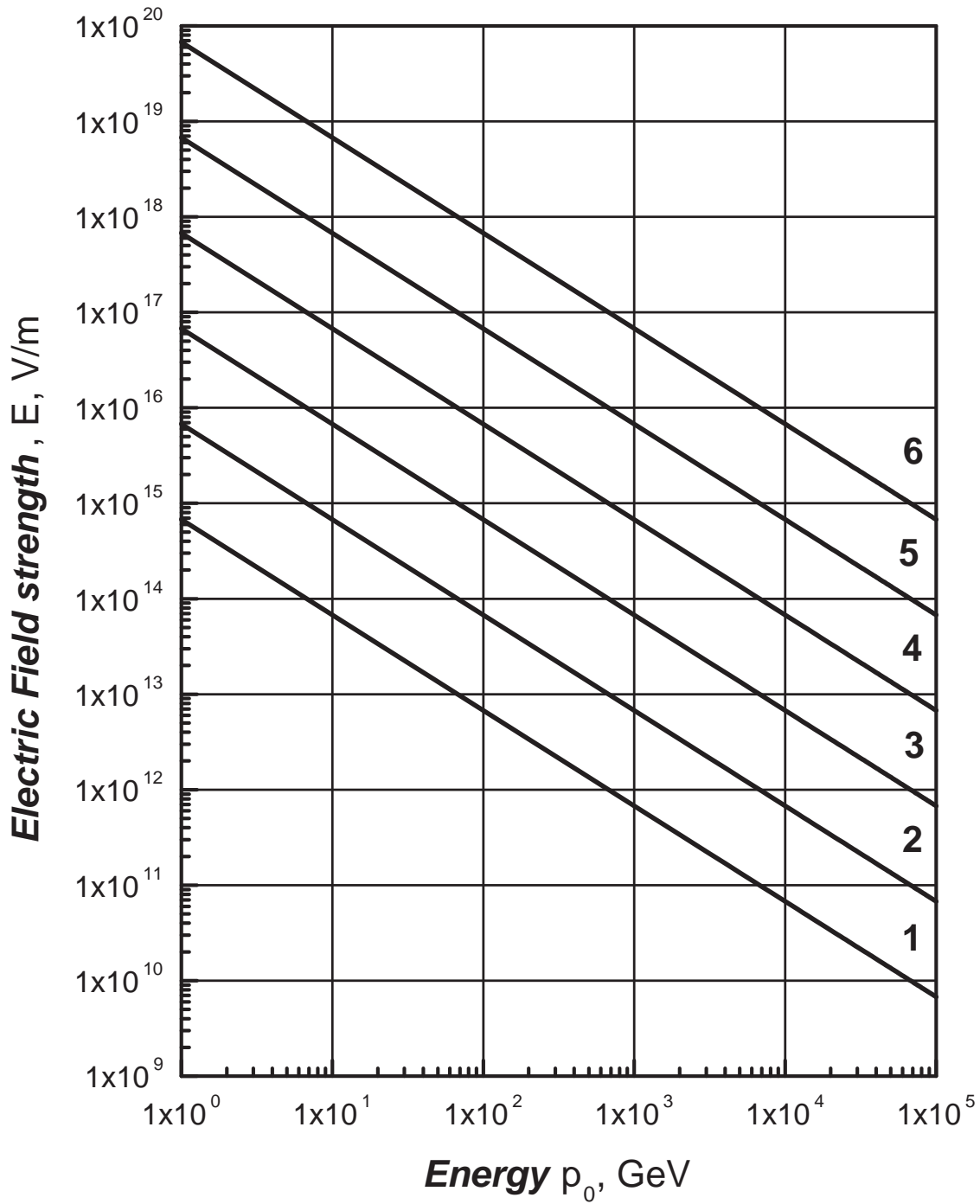}
\begin{minipage}[t]{15 cm}
\caption[]{The energy-field domains plotted for different values
of the background field parameter $\chi$ (\ref{chi}). The lines
correspond to the following choices: (1) $\chi = 1$, (2) $\chi =
10$, (3) $\chi = 100$, (4) $\chi = 10^3$, (5) $\chi = 10^4$, (6)
$\chi = 10^5$.}
\end{minipage}
\end{center}
\end{figure}

\section{Conclusion}
In this paper I have presented some methodological ideas
providing the basis for practical calculations of quantum
processes in very strong electromagnetic fields. The primary
motivation of these studies is to look for new physics that
could emerge due to the intense electromagnetic background.
There is a hope that by combining modern achievements in the
laser technique and in electromagnet construction with
traditional methods of elementary-particle physics, one could
obtain results that are inaccessible in other investigations.

Tracing some general features of quantum processes induced by
strong electromagnetic fields I have estimated the energy-field
domains that are needed in order to observe a number of reactions
which are forbidden under normal conditions. At relatively small
values of the field strength there remains the only one parameter
$\chi$ (\ref{chi}) which has a crucial influence on the rate of
the reaction and governs how the energy threshold moves with the
increase of the field intensity. I drew a figure for illustrative
purposes (see Fig.1) to display the characteristic domains
relevant to the processes mentioned in the previous section. We see
that present experimental situation is confined by the bound
$\chi\le 10$. This conclusion comes from the estimates of electric
fields extending over macroscopic distances along strings of atoms
in single crystals \cite{CERN 94-05}. For example, the electric
field along a $<111>$ axis in a crystal of Wolfram amounts to
$E\simeq 5\cdot 10^{13}\ V/m$ \cite{crystals}. For crystals of
higher atomic number, the electric fields can be still larger than
the value given above. However it is insufficient to reach the
domain $\chi\sim 10^2\div 10^5$ as long as the energies of
accelerated particles do not exceed the TeV scale. So we can expect
that with the particle machines of the new generation there could
appear a possibility for experimental observation of some
reactions discussed in this paper.

\section*{Acknowledgements}
The author would like to thank the Theory Division of CERN
for kind hospitality and financial support while part of this
work was being done.
\section*{Appendix}
We have employed the Airy functions being connected with the
ones described in mathematical handbooks in the following
manner.
\begin{equation}
\label{Airy}
\Phi(z)=\pi Ai(z), \qquad
\Phi_1(z)=\pi^2 \Bigl[Ai(z) Gi'(z)-Ai'(z) Gi(z)\Bigr], \qquad
\Phi'(z)=\pi Ai'(z)
\end{equation}
In accordance with standard mathematical notations these
functions are defined through the integral representations
\cite{Abramovitz}
\begin{equation}
\label{Airy-integral}
Ai(z)={1\over\pi}\int\limits_0^\infty dt\cos(zt+t^3/3),\qquad
Gi(z)={1\over\pi}\int\limits_0^\infty dt\sin(zt+t^3/3)
\end{equation}
and they obey the following differential equations
\begin{equation}
\label{Airy-equation}
Ai''(z)- z Ai(z)=0, \qquad Gi''(z)-z Gi(z)=-{1\over\pi}.
\end{equation}
The Airy functions that are the entire functions of the complex
variable $z$ can be expanded in the Laurent series that converge
everywhere ($\vert z \vert < \infty$)
\begin{equation}
\label{series-1}
Ai(z)={3^{-2/3}\over\pi}\sum_{n=0}^\infty
\Gamma \biggl({n+1\over 3}\biggr) \sin
\biggl( {\pi\over 3}-{2\pi n\over 3} \biggr)
\frac{(3^{1/3}z)^n}{n !},
\end{equation}
\begin{equation}
\label{series-2}
Gi(z)={3^{-2/3}\over\pi}\sum_{n=0}^\infty
\Gamma \biggl({n+1\over 3}\biggr) \cos
\biggl( {\pi\over 3}-{2\pi n\over 3} \biggr)
\frac{(3^{1/3}z)^n}{n !}.
\end{equation}
Many properties of the Airy functions can be deduced from their
interrelation with the cylindrical (Bessel and McDonald)
functions of the fractional index.
\begin{equation}
\label{Airy-Bessel-1}
Ai(z)={1\over\pi}\sqrt{z\over 3} K_{1/3}
\Bigl({2\over 3}z^{3/2}\Bigr), \qquad
Ai(-z)={\sqrt{z}\over 3}\biggl[
J_{1/3}\Bigl({2\over 3}z^{3/2}\Bigr)+
J_{-1/3}\Bigl({2\over 3}z^{3/2}\Bigr)\biggr]
\end{equation}
\begin{equation}
\label{Airy-Bessel-2}
Ai'(z)=-{z\over\pi\sqrt{3}} K_{2/3}
\Bigl({2\over 3}z^{3/2}\Bigr), \qquad
Ai'(-z)={z\over 3}\biggl[
J_{2/3}\Bigl({2\over 3}z^{3/2}\Bigr)-
J_{-2/3}\Bigl({2\over 3}z^{3/2}\Bigr)\biggr]
\end{equation}
For large values of its arguments ($\vert
z \vert\rightarrow\infty, \vert\arg z\vert <\pi$) there is
a possibility to approximate those by the following
asymptotic estimates:
\begin{equation}
\label{Airy-asymptotic_1}
Ai(z)={z^{-1/4}\over 2\pi}\exp\Bigl(-{2\over 3}z^{3/2}\Bigr)
\sum_{n=0}^\infty (-1)^n \frac{\Gamma (3n+1/2)}{(2n)!}
\bigl(9z^{3/2}\bigr)^{-n}
\end{equation}
\begin{equation}
\label{Airy-asymptotic_2}
Gi(z)={1\over\pi z}
\sum_{n=0}^\infty \frac{(3n)!}{n!} \bigl(3z^3\bigr)^{-n}
\end{equation}

\pagebreak

\end{document}